\documentclass{PoS}

\title{The LHeC Detector}

\ShortTitle{The LHeC Detector}

\author{Peter Kostka\\
  University of Liverpool, Liverpool, L69 7ZE, Great Britain\\
  Email: \email{peter.kostka@liv.ac.uk}}

\author{Alessandro Polini\\
  INFN Bologna, via Irnerio 46, 40126 Bologna, Italy\\
  E-mail: \email{alessandro.polini@bo.infn.it}}

\author{\speaker{David M. South}\thanks{on behalf of the LHeC Collaboration.}\\
  Deutsches Elektronen Synchrotron, Notkestrasse 85, 22607 Hamburg, Germany\\
  E-mail: \email{david.south@desy.de}}

\abstract{The Large Hadron Electron Collider (LHeC) is a proposed upgrade to the LHC,
  to provide high energy, high luminosity electron-proton collisions to run
  concurrently with Phase $2$ of the LHC. The baseline design of a detector for the LHeC
  is described, driven by the requirements from the projected physics programme and
  including some preliminary results from first simulations.}

\FullConference{XXI International Workshop on Deep-Inelastic Scattering and Related Subject -DIS2013,\\
		22-26 April 2013\\
		Marseilles,France}

\begin{document}

\section{Introduction to the LHeC}

The LHeC~\cite{maxDIS} is a planned $ep$ collider at CERN,
where one of the $7$~TeV LHC proton beams is brought into collision with a $60$~GeV
electron\footnote{The term ``electron'' is used generically to refer to both electrons and positrons, unless
otherwise stated.} beam, operating with a design luminosity of about $10^{33}$~cm$^{-2}$~s$^{-1}$.
Data taking at the LHeC is to run simultaneously to the main LHC $pp$ collisions
during Phase $2$ operation (beyond 2023).
Two alternative accelerator designs are considered: a ring-ring approach similar to that employed at HERA,
requiring the installation of a lepton ring in the LHC tunnel and a linac-ring design employing two $1$km
long linacs, with energy recovery~\cite{lheccdr}.
The linac-ring design is the favoured option due to the reduced impact on the LHC schedule and the interest
in the machine physics community of the novel new accelerator design~\cite{oliverDIS}.


The LHeC combines high energy, high precision and high luminosity in a wide reaching
physics programme to complement that of the LHC.
Not only does the LHeC greatly extend the reach of precision DIS physics~\cite{mandyEPS} when
compared to results from fixed target experiments or from HERA, but opens up an area of low-x phase space
not previously explored~\cite{nestorDIS}.
The LHeC enables aspects of the photoproduction of massive gauge bosons to be studied in a new
kinematic window~\cite{magnoDIS}.
The determination of proton and nuclear PDFs~\cite{hannuDIS} will also form part of the core physics
programme, in addition to measurements in the top sector.
The LHC will run with ion beams during Phase $2$, providing the possibility of an $eA$ dataset at
the LHeC~\cite{maxEPS}.
A substantial range of searches for new physics can be undertaken at the LHeC, including
among others leptoquarks, RPV SUSY and contact interactions~\cite{newphysEPS}.
Finally, the production of Higgs bosons via vector-boson-fusion in charged current with
decay to $b\bar{b}$ may be examined at the LHeC, where the final state produces a much cleaner
signal than at the LHC~\cite{bruceDIS}.
This is particularly interesting if the latest estimates for the LHeC luminosity are realised, where the
design luminosity is exceeded by a factor of $10$ and the LHeC can truly be considered a Higgs facility~\cite{utaEPS}.


These proceeedings present the baseline design of the proposed LHeC detector.
The basic detector requirements are detailed in section~\ref{sec:design}, followed by a
description of the main detector in section~\ref{sec:maindet}.
Additional detectors to be placed in the tunnel are briefly described in section~\ref{sec:tunnel}. 
%
An outlook is given in section~\ref{sec:outlook}.

\section{Detector requirements}
\label{sec:design}

The LHeC detector needs to be designed, constructed and ready for use at the beginning of LHC Phase $2$,
which is approximately $12$ years from now.
Such a timescale prohibits a dedicated, large scale R\&D programme, but the LHeC detector can profit
from current and upgrade LHC technologies, as well as ILC development, and the wealth of
experience gained in $ep$ detectors at HERA.
It must be able to run concurrently with the other LHC $pp$ and $pA$ experiments, in order to make
use of the LHC beam and record the corresponding $ep$ and $eA$ data.
The detector would be located at Point~$2$, the only interaction point of the LHeC beams.
The LHeC interaction region (IR) provides additional constraints from the necessarily complex optics
in handling $3$ beams: the electron and proton beams colliding head on, as well as the second,
non-colliding LHC proton beam~\cite{lheccdr}, which must pass through unimpeded.
Furthermore a long dipole system (-$9$m$<z<+9$m) of $0.3$ T is required across the whole interaction
region to steer the electron beam in order to achieve the design luminosity, by reducing the
crossing angle and separating the beams after the interaction.
The LHeC detector should also be modular and flexible in design, with assembly above ground where
possible, in order to accommodate future upgrades.
Finally, it should be affordable, with a reasonable cost compared to other similarly built detectors.
In order to fulfil all facets of the experimental programme, the following physics requirements are desirable:

\vspace{-0.3cm}

\begin{itemize}

\item{A high resolution tracking system to provide excellent primary vertex resolution and resolution of
    secondary vertices down to small angles in forward direction for high $x$ heavy flavour physics and
    searches. A precise $P_{T}$ measurement, matched to calorimeter signals calibrated and aligned to an
    accuracy of  $1$~mrad.}

\vspace{-0.3cm}
  
\item{Full coverage calorimetry. Electron energy measured to $10\% / \sqrt{E}$, calibrated using
    the kinematic peak and double angle method to the per-mil level. Hadronic energy measured to
    $40\% / \sqrt{E}$, with a calibrated $P_{T}$ balance to an accuracy of $1$\% .}
  
\vspace{-0.3cm}

\item{Tagging of backward scattered photons and electrons for a precise measurement of luminosity and
    photoproduction physics. Tagging of forward scattered protons, neutrons and deuterons to fully investigate
    diffractive and deuteron physics.}  
  
\vspace{-0.3cm}

\item{A muon system, for tagging and momentum measurement in combination with tracking}.
 
\end{itemize}

\vspace{-0.3cm}

The remainder of these proceedings outlines the detector configuration and the baseline design chosen,
as described in the LHeC Conceptual Design Report~\cite{lheccdr}.

\section{Central detector}
\label{sec:maindet}

\begin{figure}[h]
  \begin{center}
    \includegraphics[width=0.75\textwidth]{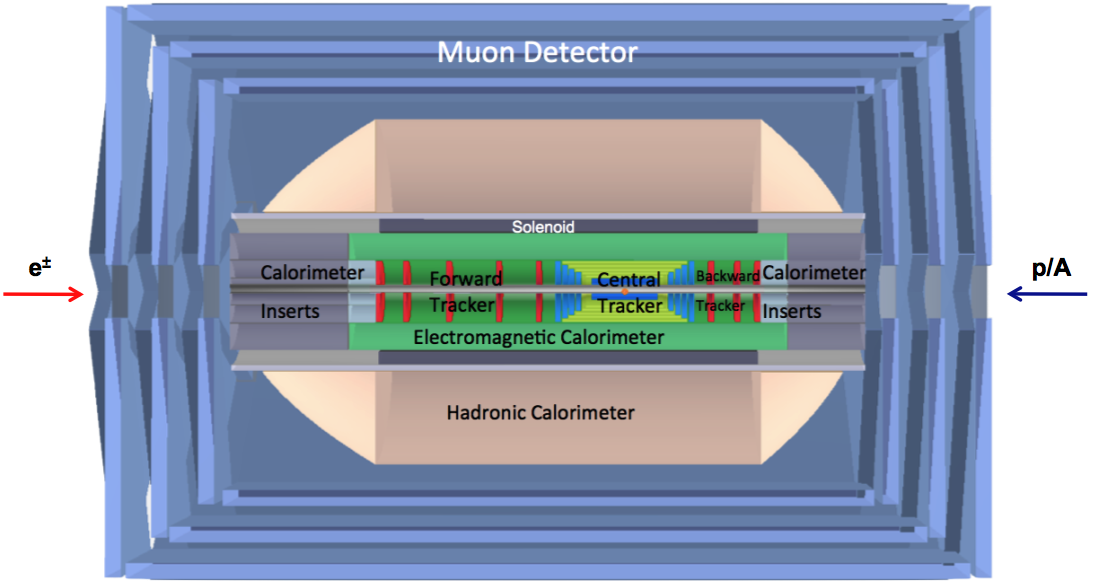}
  \end{center}
  \vspace{-0.4cm}
  \caption{An $r$-$z$ cross section of the main LHeC detector in the baseline configuration with the solenoid
    and dipoles placed between the EMC and the HAC. The proton beam, from the right, collides with the
    electron beam, from the left, at the IP which is surrounded by a central tracker system complemented
    by larger forward and backward tracking detectors and followed by sets of calorimeters.
    On the outside, a muon system completes the picture.}
 \label{fig:centraldet}
  \vspace{-0.2cm}
\end{figure}

The LHeC detector design has two main and necessary features: hermetic coverage,
especially in the forward and backward directions to provide a precise energy measurement
and an intrinsic asymmetry, reflecting the corresponding asymmetry in the beam energies.
Both of these features can be seen in figure~\ref{fig:centraldet}, which illustrates the central, main
part of the proposed LHeC detector.
The dimensions of the main LHeC detector are $14$m $\times 9$m, small enough to fit inside
the L3 magnet at Point~$2$, and much smaller than the CMS ($21$m $\times 15$m) or ATLAS
($45$m $\times 25$m) detectors.
The key components of the central detector are described below.

\subsection{Tracking}
\label{sec:tracking}

\begin{figure}[h]
  \begin{center}
    \includegraphics[width=0.85\textwidth]{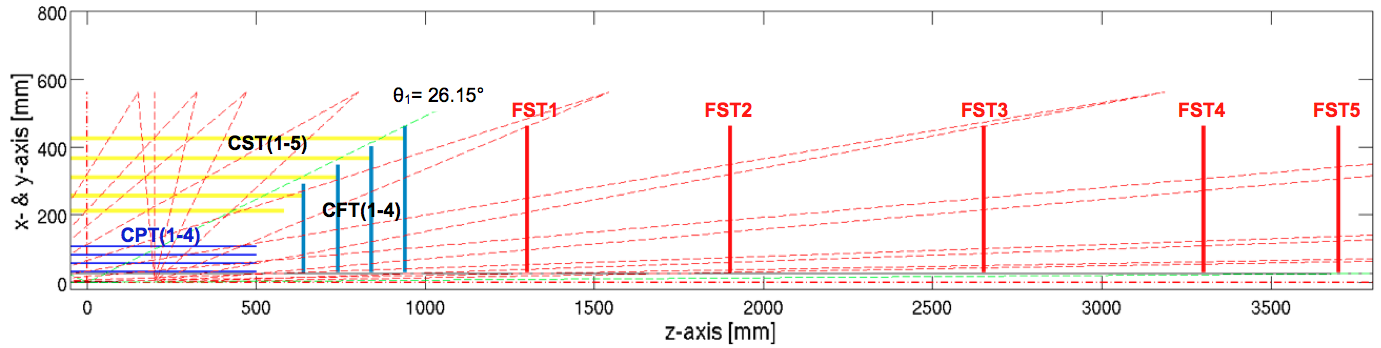}
  \end{center}
  \vspace{-0.4cm}
  \caption{The LHeC tracking design of the central and forward region, simulated using LicToy2.0 program.}
 \label{fig:lictoy}
  \vspace{-0.2cm}
\end{figure}

The LHeC detector has a high acceptance, compact tracking design, completely contained
within the electromagnetic calorimeter (EMC).
An all silicon design is employed, using pixel and strip detectors, with more
coverage in the proton direction and an elliptical arrangement of those layers closest to
the similarly shaped beampipe.
Silicon is chosen because of its compact design, low material budget, and radiation hardness,
although dedicated studies of neutron fluences with GEANT4~\cite{geant4} and FLUKA~\cite{fluka}
show that the LHeC environment is not as challenging as at the LHC~\cite{lheccdr}.
The tracking design of the central and forward region of the LHeC is shown in
figure~\ref{fig:lictoy}, as simluated using LicToy~\cite{lictoy}.
Transverse momentum resolutions down to $10^{-3}~{\rm GeV}^{-1}$ and
impact parameter resolutions of distances as small as $10~\mu m$ are expected from the simulation.

\subsection{Calorimetry and magnets}
\label{sec:calo}

\begin{figure}[h]
  \begin{center}
    \includegraphics[width=0.3\textwidth]{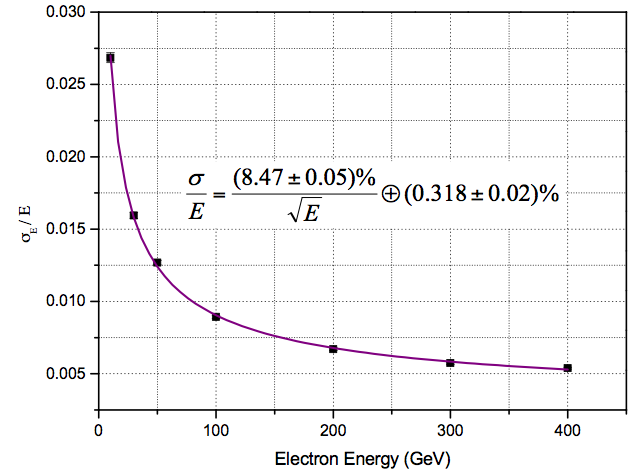}
    \hspace{0.1cm}
    \includegraphics[width=0.61\textwidth]{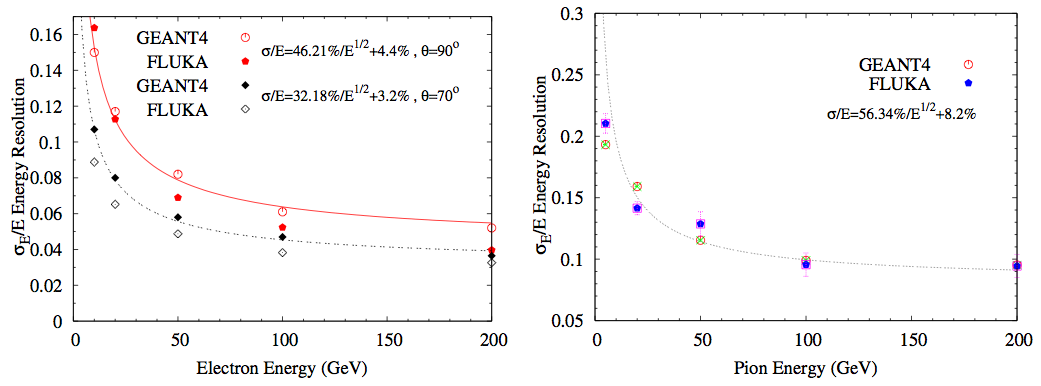}
  \end{center}
  \vspace{-0.4cm}
  \caption{Left: The EMC energy resolution for electrons between $10$ and $400$~GeV
    (simulated using GEANT4). Middle and right: A comparion of HAC energy resolution simulated
    using GEANT4 and FLUKA for electrons at $\theta_e$ at $70^{\circ}$ and $ 90^{\circ}$ (middle) and pions at
    $\theta_{\pi}$ at $90^{\circ}$ (right).}
\label{fig:calostudies}
  \vspace{-0.2cm}
\end{figure}

The main EMC in the barrel region, $2.8 < \eta< -2.3$, is based on the LAr/Pb design used in ATLAS
and employs three different granularity sections longitudinally with radiation lengths $25-30~X_{0}$.
The baseline hadronic calorimeter (HAC) design uses steel absorber and scintillator sampling plates,
similar to the TILE calorimeter in ATLAS, with interaction lengths $7-9~\lambda_{i}$.
The calorimeter structure provides support for inner detectors and return flux for the solenoid.
Complementing the main calorimeters are electromagnetic and hadronic inserts in both the forward
and backward regions.
The baseline magnet design is a small $3.5$T coil between EMC and HAC,
placing the solenoid and dipoles conveniently within the same cold vacuum vessel.
Many simulation studies have been performed with GEANT4 and FLUKA to investigate properties such as 
containment, resolution, combined response, and the dependence on the position of solenoid, dipoles and cryostat.
Simluated energy resolutions are shown in figure~\ref{fig:calostudies}; more details are in the CDR~\cite{lheccdr}.

\subsection{Muon system}
\label{sec:muon}

The baseline muon system consists of $2$-$3$ layers, each with a double trigger layer and a layer for 
measurements, employing standard, commonly used technologies such as thin gap chambers, resistive
plate chambers and drift tubes.
In this design, the muon system provides no independent measurement and the muon momentum
is measured by the inner tracker, in combination with signals from the muon system.
Several muon system extensions are possible, including providing an independent momentum measurement,
a larger solenoid or dual coil system (with all of calorimeter within inner coil) and a forward toroid (air core design)~\cite{lheccdr}.

\section{Tunnel systems}
\label{sec:tunnel}

\begin{figure}[h]
  \begin{center}
    \includegraphics[width=0.52\textwidth]{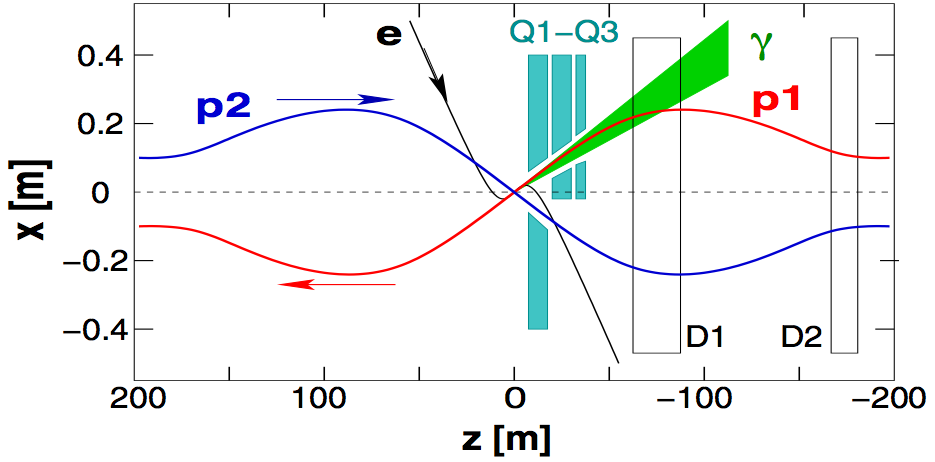}
    \hspace{0.2cm}
    \includegraphics[width=0.44\textwidth]{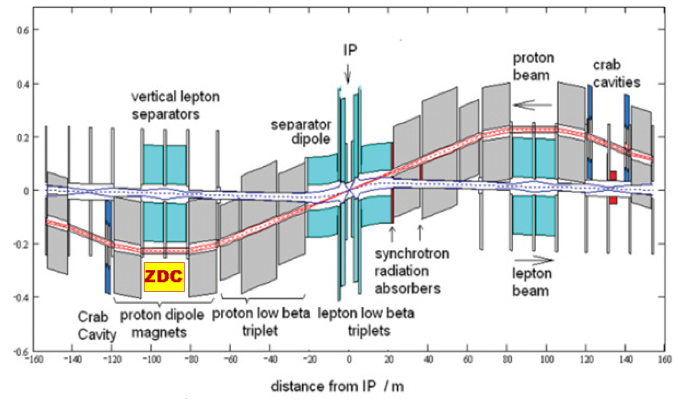}
  \end{center}
  \vspace{-0.4cm}
  \caption{Left: Schematic view of the LHeC IR for the linac-ring option, showing
    the colliding proton beam ($p1$), the non-colliding proton beam ($p2$), the electron beam ($e$)
    and a $3\sigma$ fan of Bethe-Heitler photons ($\gamma$) used for the luminosity measurement.
    Right: Another view of the LHeC IR with a similar scale, showing the possible position of the ZDC.}
  \label{fig:tunnelsystems}
  \vspace{-0.2cm}
\end{figure}

In the linac-ring LHeC configuration, Bethe-Heitler (BH) collinear photons travel
along with the proton beam and are detected at $z \approx -120$~m, after
the proton bending dipole as illustrated in figure~\ref{fig:tunnelsystems} (left).
Whereas BH photons can be used to measure the luminosity, photons from QED Compton
events may also be used, where the visible cross section for such processes can be
enhanced by the introduction of an additional QEDC tagger at $z \approx -6$m.
An electron-tagger may also be installed to detect the scattered electron in
BH events, benefiting photoproduction analyses, where the
preferred distance from acceptance arguments is $z \approx -62$~m.

In the forward direction, the Zero Degree Calorimeter (ZDC) may be used to measure
the energy and angles of very forward particles in tunnel $z \approx +100$~m, as
illustrated in figure~\ref{fig:tunnelsystems} (right).
Such a device operates in a very demanding radiation environment and features a
Tungsten-Quartz design, and the exact position and detector dimensions depend on the
space available for installation, to be determined by simulation.
Even further down the beamline in the proton direction, at $z \approx +420$~m,
an additional proton detector may enhance the diffractive physics programme at LHeC,
where relevant R\&D detector studies have already been performed at the LHC.

\section{Outlook}
\label{sec:outlook}

A LHeC baseline detector concept has been worked out, as described in the CDR~\cite{lheccdr}.
The design depends heavily on the constraints from the machine, the interaction region and the
LHC activities.
A feasible and affordable concept, fulfilling the physics requirements has been presented.
With respect to the baseline many improvements may become available; a more precise 
design will follow from more detailed simulations, engineering and knowledge of machine 
constraints.

\end{document}